\documentclass[pra,aps,showpacs,twocolumn,superscriptaddress]{revtex4}
\usepackage{amsmath}
\usepackage{graphicx}
\usepackage{subfigure}

\begin{document}

\title{Different time scales in plasmonically enhanced high-order harmonic generation}
\author{C. Zagoya, M. Bonner, H. Chomet, E. Slade and C. Figueira de Morisson Faria\\ Department of Physics and Astronomy, University College London, Gower Street, London WC1E 6BT, United Kingdom}
\date{\today}

\begin{abstract}
We investigate high-order harmonic generation in inhomogeneous media for reduced dimensionality models. We perform a phase-space analysis, in which we identify specific features caused by the field inhomogeneity. We compute high-order harmonic spectra using the numerical solution of the time-dependent Schr\"odinger equation, and provide an interpretation in terms of classical electron trajectories. We show that the dynamics of the system can be described by the interplay of high-frequency and slow-frequency oscillations, which are given by Mathieu's equations. The latter oscillations lead to an increase in the cutoff energy, and, for small values of the inhomogeneity parameter, take place over many driving-field cycles. In this case, the two processes can be decoupled and the oscillations can be described analytically.
\end{abstract}
\maketitle

\section{Introduction}
\label{sec:intro}

High-order harmonic generation (HHG) is a nonlinear phenomenon in which a high-intensity, low frequency field focused on a gaseous target leads to extreme ultraviolet (XUV) radiation. Since its first observation, in the late 1980s, it has led to a myriad of applications, such as attosecond imaging of matter, subfemtosecond spectroscopy and XUV sources \cite{Krausz_2009}. They are based on the fact that HHG is caused by the laser-induced recombination of an electron with a bound state of its parent ion. Thereby, the three main steps the active electron must undertake are (i) tunnel ionization, (ii) propagation in the continuum, in which the electron is accelerated by the field, and (iii) recombination, in which the electron's kinetic energy is released in form of high-order harmonics \cite{Corkum_1993,Lewenstein_1994}.  

Typical features in high-order harmonic spectra are a broad energy region with harmonics of comparable intensities, the so-called plateau, followed by a sharp decrease in the harmonic yield, the so-called cutoff. These features are very favorable for the above-stated applications, in particular XUV sources. Unfortunately, however, the intensity of high-order harmonics is several orders of magnitude lower than that of the fundamental. Hence, throughout the years, a major challenge has not only been to increase the cutoff frequency, but additionally, the  HHG efficiency. 

With those two aims in mind, alternative media have attracted a great deal of interest in recent years as potential HHG sources.  
For instance, it has been reported that a laser field enhanced by plasmonic resonances in the vicinity of nanostructures gave rise to high-order harmonics \cite{Kim_2008}. Although there has been some controversy about these results \cite{Sivis_2011,Kim_2011,Sivis_2013}, such media, as well as metal nanotips \cite{Schenk_2010,Kruger_2011}, dielectric nanospheres \cite{Zherebtsov_2011} and ablation plumes \cite{Hutchison_12,Ganeev_2012} are highly promising harmonic sources, and allow a greater deal of control than traditional, gaseous media as they can be engineered for very specific purposes. 

An important feature in the above-mentioned sources is that the external driving field exhibits a high degree of spatial confinement and can no longer be regarded as spatially homogeneous \cite{Thongrattanasiri_2013}. 
Recent theoretical investigations addressed this issue by considering a simplified one-electron model in which the driving field has been made spatially dependent \cite{Husakou_2011,ciappina2012,Ciappina2012b,shaaran2012,Yavuz_2012,shaaran2013,Hekmatara_2014,Luo_2014}. Therein, it has been shown that the inhomogeneity leads to a significant extension in the cutoff energy. This increase has been related to a higher kinetic energy for the returning electron, which has been identified in the inhomogeneous case. The symmetry breaking for subsequent half cycles that occurs for inhomogeneous fields, together with the increase in the electron momentum upon ionization, have been reported as causes for the cutoff extension \cite{shaaran2012}. Therein, it has been also shown that, for larger values of the inhomogeneity parameter, the electron's return along the long orbit in the dominant pair is suppressed. Furthermore, extremely high electron return energies have been identified for longer pairs of orbits. They have 
been attributed to the electron spending a long time in the continuum. 

One should note, however, that, although the above-mentioned features have been investigated in detail, these studies so far have remained at the descriptive level. In fact, it is not clear why the electron's return via the long orbit is hindered by the field inhomogeneity. Furthermore, a longer time in the continuum does not necessarily lead to a higher return energy for the active electron. For instance, if the driving field is homogeneous, the shortest pair of orbits will lead in many cases to a higher returning energy than the longer pairs \footnote{Specifically for a monochromatic field, the first return gives a kinetic energy of $3.17U_p$, where $U_p$ is the ponderomotive energy, while longer returns give lower energies. }. Finally, a higher momentum  upon ionization is not the sole mechanism leading to an increase in the cutoff energy. For instance, an additional confining potential may force the electron to return to the core along high-energy orbits, instead of ionizing irreversibly \cite{FR_2000}. 
Hence, the above-stated cutoff increase, the suppression of the long orbit, and the symmetry breaking deserve a closer look.

In this work, we address these issues by analyzing the electron dynamics in phase space.
Often, phase-space considerations provide additional insight on aspects that are commonly overlooked. Examples are the interplay between the electric field and the binding potential in NSDI with circularly polarized driving fields \cite{Mauger_2010,Kamor_2013}, and  the relation between HHG and closed orbit theory \cite{Mauger_2014}. We compute HHG spectra for one-dimensional models, in which the electronic wavepacket propagation is calculated using the time-dependent Schr\"odinger equation. The electron return times are extracted from the time-dependent dipole acceleration using windowed Fourier transforms, which are compared with classical-trajectory computations. We also provide an analytical model valid for small inhomogeneity parameters based on Mathieu's equation and the Dehmelt approximation, which relates the features encountered to different time scales.

This article is organized as follows. In Sec.~\ref{sec:backgd}, we give the necessary theoretical background in order to understand the subsequent results. These are presented in Sec.~\ref{sec:results}, in which we discuss the phase-space features encountered for homogeneous and inhomogeneous fields (Sec.~\ref{subsec:phasespace}), analyze the HHG spectra in terms of classical trajectories (Sec.~\ref{subsec:classical}) and provide an analytical model for the kinetic energy increase (Sec.~\ref{subsec:analytical}). Finally, in Sec. \ref{sec:conclusions} we state our main conclusions. We use atomic units throughout.
\section{Background}
\label{sec:backgd}
\subsection{Model}
The Hamiltonian associated with the model-atom in a inhomogeneous medium is given by 
\begin{equation}
H=\frac{p^2}{2}+V_a+V_l, \label{eq:Hamiltonian}
\end{equation}
where $V_a$ denotes the atomic potential
and 
\begin{equation}
V_l(x,t)=xE(x,t) \label{eq:vlaser}
\end{equation}
gives the interaction with the driving field. We use $x$ and $p$ to represent the position and momentum of the electron, respectively. Here we incorporate the inhomogeneity in a similar way as in \cite{shaaran2012}, which is a good approximation for the field generated near metal nanospheres as long as the inhomogeneity parameter is small \cite{shaaran2013,Sussman_2011,Sussman_2011b}.
This yields
\begin{equation}
E(x,t)=(1+\beta x){\cal E}_t, \label{eq:generalfield}
\end{equation}
where 
\begin{equation}
{\cal E}_t=E_0f(t)\cos(\omega t +\phi)\label{eq:tdfield}
\end{equation}
denotes the time-dependent part of the field and $\beta$ gives the inhomogeneity parameter.

 The field amplitude and frequency are given by $E_0$ and $\omega$, respectively, while $\phi$ is an arbitrary phase. In our computations, we use a flat-top pulse, which means that $f(t)=1$ for $0\le t \le t_f$, where $t_f$ is the pulse duration, and zero otherwise. 

We choose $V_{a}$ to be the soft-core potential
\begin{equation}
V_{a}(x)=-\frac{1}{\sqrt{x^2+1}}.\label{eq:softcore}
\end{equation}
The initial wave packet is chosen to be of the form
\begin{equation}
\Psi(x,0)=\left(\frac{\gamma}{\pi}\right)^{1/4}\exp\left[-\frac{\gamma}{2}(x-q_0)^2+\mathrm{i}p_0(x-x_0)\right],\label{eq:psi0}
\end{equation}	
where $x_0$ and momentum $p_0$ refer to the initial coordinate and momentum, respectively. The width of the initial wave packet is kept fixed at $\gamma=0.5$, which gives a ground-state energy of $-I_p=-0.67$ a.u. This wavepacket is then propagated in time, and  $\Psi(x,t)$ is computed solving the time-dependent Schr\"odinger equation (TDSE)
\begin{equation}
i \frac{\partial \Psi(x,t)}{\partial t} = H \Psi(x,t)
\label{eq:schro}
\end{equation}
numerically with the split-operator method.

We compute the high-order harmonic spectrum using the dipole acceleration \cite{Sundaram1990,Burnett_1992,Krause_1992}
\begin{eqnarray}
a(t)=&-&\langle\Psi(t)|\frac{\partial V_{\mathrm{eff}}(x,t)}{\partial x}|\Psi(t)\rangle \notag \\
=&-&\int \Psi^*(x,t)\frac{\partial V_{\mathrm{eff}}(x,t)}{\partial x}\Psi(x,t)\mathrm{d}x,
\label{eq:acceleration}
\end{eqnarray}
where $V_{\mathrm{eff}}(x,t)$ is the effective potential given by
\begin{equation}
V_{\mathrm{eff}}(x,t)=V_l(x)+V_a(x).\label{eq:effective}
\end{equation}
\subsection{Fourier and Gabor spectra}
The HHG spectrum is computed as $\chi(\Omega)=|a(\Omega)|^2$, with
\begin{equation}
a(\Omega)=\int\mathrm{d}t\,a(t)\mathrm{e}^{-\mathrm{i}\Omega t}.
\end{equation}
In order to compute time-resolved spectra, we employ $\chi_G(\Omega,t)=|a_G(\Omega,t)|^2$, where 
\begin{equation}
a_{\mathrm{G}}(\Omega,t)=\int\mathrm{d}t'\,a(t')\mathrm{e}^{-\mathrm{i}\Omega t'-(t'-t)^2/2\sigma^2}\, 
\label{eq:Gabor}
\end{equation}
with $\sigma=1/3\omega$ is a windowed Fourier transform with a Gaussian window function. Eq.~(\ref{eq:Gabor}) is known as the Gabor transform, and is a well-established method for extracting the electron return times from the TDSE spectrum (see, e.g., Refs.~\cite{Antoine_1995,FDS1997,deBohan_1998,FDS1999,Tong_2000} or  Refs.~\cite{Chirila_2010,ciappina2012,Ciappina2012b,Wu_2013_2}  for early studies or more recent articles, respectively). The limit $\sigma \rightarrow \infty$ gives the standard Fourier transform, for which all temporal information is lost.

\section{results}
\label{sec:results}
\subsection{Phase-space regions and main features}
\label{subsec:phasespace}

\begin{figure*}
	\hspace*{-0.8cm}\includegraphics[scale=0.6]{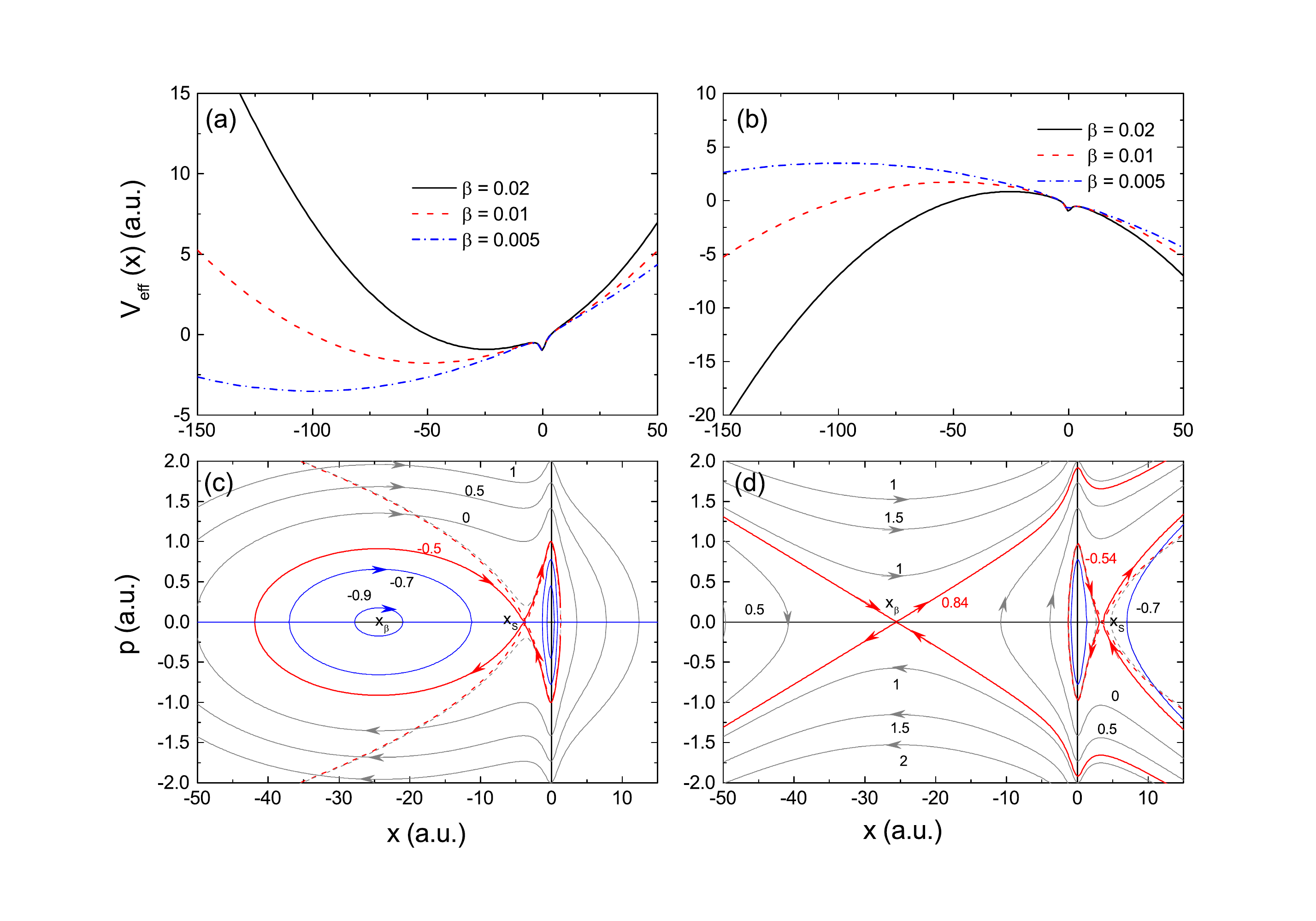}
	\caption{\label{fig:phaseportraits} Effective potentials $V_{\mathrm{eff}}(x,t)$ for spatially inhomogeneous fields, considering several values of $\beta$, 
		 ${\cal E}_t=0.07$ a.u. and ${\cal E}_t=-0.07$ a.u. [panels (a) and (b) respectively], together with the phase portraits for the Hamiltonian (\ref{eq:Hamiltonian}) and $\beta=0.02$ [lower panels]. In panels  (c) and (d), we consider the field to be  ${\cal E}_t=0.07$a.u., and  ${\cal E}_t=-0.07$ a.u., respectively.
		   The separatrices are given by the red lines in the figure, and the numbers near each contour denote the corresponding total energy of the system. The Stark saddle $x_S$ and the fixed point $x_{\beta}$ due to the inhomogeneity are indicated in the figure. The contours in blue are related to the energies lower than that of the Stark saddle. The red dashed lines give the separatrices for the homogeneous case $\beta=0$, which occur at at energy $E_{\mathrm{sep}}=-0.52$ a.u. The black dashed lines give the phase-space trajectory for $E=-0.5$ a.u. and $\beta=0$.}
\end{figure*}
We will first assess how the field inhomogeneity influences the effective potential $V_{\mathrm{eff}}(x,t)$ in Eq.~(\ref{eq:effective}). In general, the  laser potential (\ref{eq:vlaser}) may be written as
\begin{equation} \label{eq:HOpotential}
V_l(x,t)={\cal E}_t\beta\Big(x + \frac{1}{2\beta}\Big)^2 -\frac{{\cal E}_t}{4\beta}.
\end{equation}

According to the expression chosen for ${\cal E}_t$, there are two possibilities within a field cycle. If ${\cal E}_t>0$, the inhomogeneity introduces a concavity in the effective potential barrier, so that Eq.~(\ref{eq:HOpotential}) corresponds to a simple harmonic oscillator centered at $x_{\beta}=-1/2\beta$ a.u. and shifted by an energy $-{\cal E}_t/4\beta$ a.u. From Eq.~(\ref{eq:HOpotential}) we see that the minimum of $V_l(x)$ and its ground-state energy becomes deeper for decreasing $\beta$. 
If, in contrast,  ${\cal E}_t<0$, the inhomogeneity will render the effective barrier convex, i.e., ${\cal E}_t\beta<0$, and there will be an additional maximum for $V_{\mathrm{eff}}(x,t)$ at approximately $x_{\beta}=-1/2\beta$. For clarity, these two configurations are presented in the upper panels of Fig.~\ref{fig:phaseportraits}.

More detail about these two configurations is provided by the phase-space dynamics of the system. These dynamics are described by the differential equations
\begin{align}
\label{eq:pssystem}
\dot x&=p\\ \nonumber
\dot p&=-\frac{\partial V_{\mathrm{eff}}}{\partial x}.
\end{align}
The critical points of the system defined by (\ref{eq:pssystem}) are the points $(x_c,p_c)$ for which $\dot x=0$ and $\dot p=0$. The former condition implies that $p_c=0$, while the latter gives   $\partial V_{\mathrm{eff}}/\partial x=0$ at $x=x_c$. 
The maxima and minima of $V_{\mathrm{eff}}(x,t)$ will give the fixed points of the system, whose nature and number will change with the instantaneous field $-E_0\leq{\cal E}_t\leq E_0$. For nonvanishing field ${\cal E}_t$, they are three in total and located at $\mathbf{x}_1\simeq(x_{\beta},0)$, $\mathbf{x}_2=(x_S,0)$ and at $\mathbf{x}_3\simeq (0,0)$, where $x_{\beta}$ is defined above and $x_S$ is the Stark saddle.  The Stark saddle stems from the atomic potential $V_a(x)$ being distorted by the interaction $V_l(x,t)$, and will switch sides depending on whether ${\cal E}_t$ is positive or negative.  The fixed point near the origin is related to the minimum of $V_a(x)$.

The corresponding phase portraits for ${\cal E}_t>0$ and ${\cal E}_t<0$ are given in the  panels (c) and (d) of Fig.~\ref{fig:phaseportraits}, respectively. For ${\cal E}_t>0$ [Fig.~\ref{fig:phaseportraits}(d)], $\mathbf{x}_1$ will be a center, and the Stark saddle will be located between this fixed point and that near the origin. This means that an electronic wave packet initially localized at the origin would leave the atom by tunnel or over the barrier ionization towards negative values of $x$ and could be, in principle, trapped near $x_{\beta}$. This is very likely to occur for small values of $\beta$, as in this case the energy of this center is much lower than that of the atomic ground state [Fig.~\ref{fig:phaseportraits}(a)].

For the other half cycle (${\cal E}_t<0$), the fixed point caused by the inhomogeneity will be a saddle, and the Stark saddle will occur at $x_S>0$. The center near the origin will be located between both saddles [Fig.~\ref{fig:phaseportraits}(c)]. We expect ionization to occur for the positive values of $x$ via the Stark saddle, as the energy of the saddle at $x_{\beta}$ is much higher. This additional saddle may however function as a barrier for events that started at a previous half cycle of the field, by preventing their return to the origin. 

We will now discuss how the inhomogeneity of the field influence the Stark saddles, and thus the ionization of the electronic wave packet. If the field is homogeneous, the saddles are symmetric with regard to the origin for subsequent half cycles of the driving field. For a soft-core potential, $x_S\simeq \pm 1/\sqrt{|{\cal E}_t|}$, so that $V_{\mathrm{eff}}(x_S,t)\simeq -2\sqrt{|{\cal E}_t|}$. This is the energy of the dashed lines in Figs.~\ref{fig:phaseportraits}(c) and (d), which give the separatrices for the homogeneous field.  Upon ionization, the electronic wave packet will follow this separatrix very closely. If the absolute value of the electron momentum is lower or higher than that of the separatrix, there will be tunnel or over-the-barrier ionization, respectively. This has been discussed in our previous work \cite{Zagoya_2014}, using Wigner quasiprobability distributions. Therein, we have shown that the Wigner function exhibits a tail, which follows the separatrix very closely. This tail can be 
associated with the part of the wave packet that is freed in the continuum (see also \cite{Czirjak_2000,Graefe_2012}).

If $\beta \neq 0$, the Stark saddles are no longer symmetric for subsequent half cycles of the field. For ${\cal E}_t>0$, there will be an increase in the energy of the separatrix and a decrease in its slope near this point [see solid red line in Fig.~\ref{fig:phaseportraits}(c)], while for ${\cal E}_t<0$ there will be a decrease in the energy of the saddle and an increase in the slope, as shown in Fig.~\ref{fig:phaseportraits}(d). Hence, the minimal energy and the absolute value of the momentum with which the electron will reach the continuum will differ upon half a cycle of the field. We have verified, using Wigner probability distributions, that this is indeed the case for inhomogeneous fields, as the tail of the Wigner function associated with ionization closely follows the separatrices (not shown).
\begin{figure}
\hspace*{-0.8cm}\includegraphics[scale=0.35]{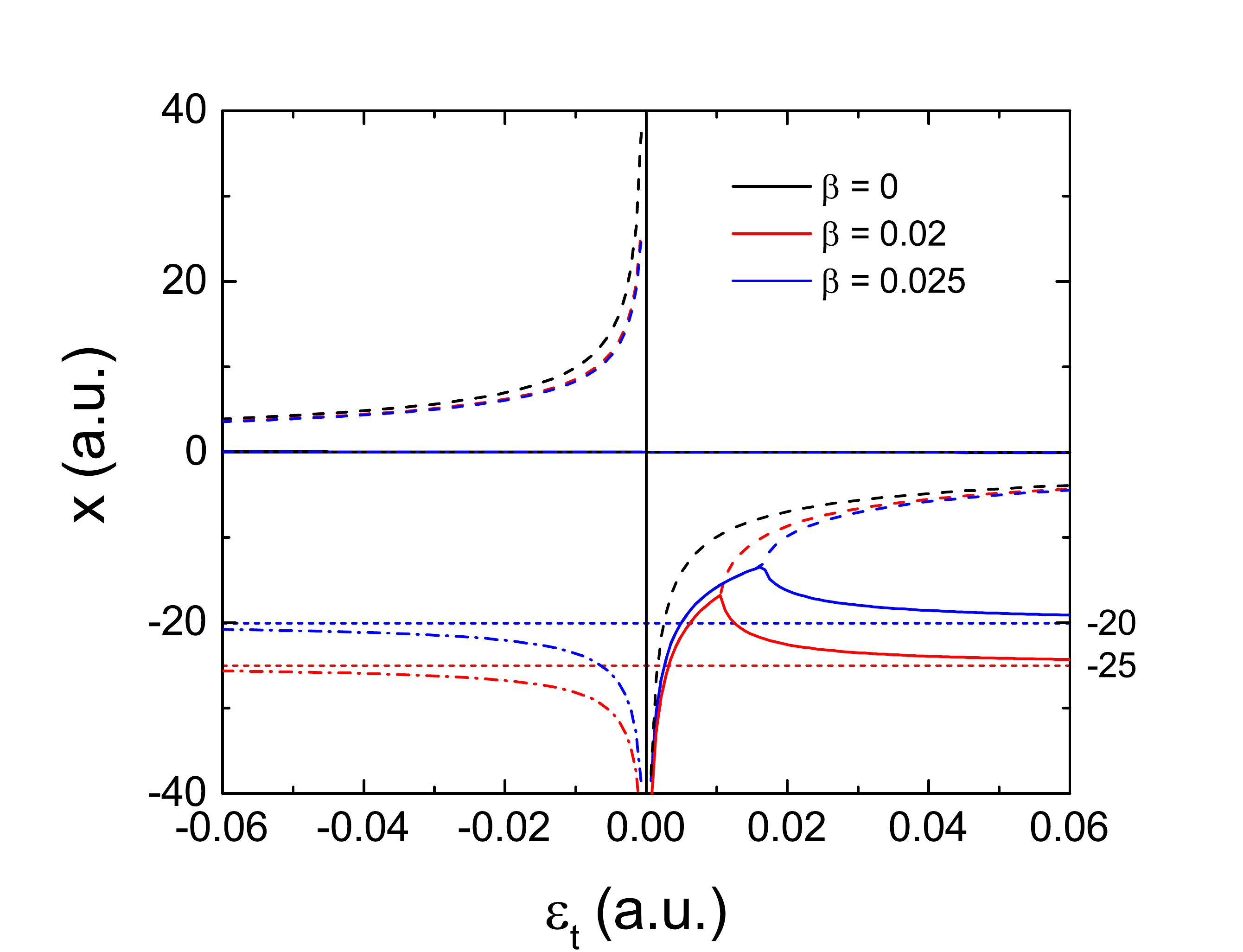}
\caption{\label{fig:eqcurve} Equilibrium curves as functions of the instantaneous field value ${\cal E}_t$, for different values of the inhomogeneity parameter $\beta$. The dashed  lines correspond to the Stark saddle, while the dot-dashed and solid lines give the saddle and center caused by the inhomogeneity, respectively. The fine short-dashed lines give the asymptotic value $x_{\beta}=-1/(2\beta)$, which will be reached for ${\cal E}_t\rightarrow \infty$. }
\end{figure}

Fig.~\ref{fig:eqcurve} shows in more detail how the three fixed points behave with regard to ${\cal E}_t$ and $\beta$. For all cases, there is a center near the origin, which will vary very little with the instantaneous field. The other fixed points will occur at $x \rightarrow \pm \infty$ for ${\cal E}_t\rightarrow 0$.  As the field increases, the Stark saddle will tend to vanishingly small $x_S$, while the fixed point caused by the inhomogeneity will tend asymptotically to $x_{\beta}=-1/(2\beta)$. 
 For ${\cal E}_t>0$ it can happen that the Stark saddle and this fixed point merge. By computing the maxima of $V_{\mathrm{eff}}(x,t)$, one may show that this occurs at ${\cal E}_t=27\beta^2$. 

\subsection{Spectra and classical-trajectory analysis}
\label{subsec:classical}
\begin{figure}
	\centering
	\includegraphics[width=8cm,angle=0]{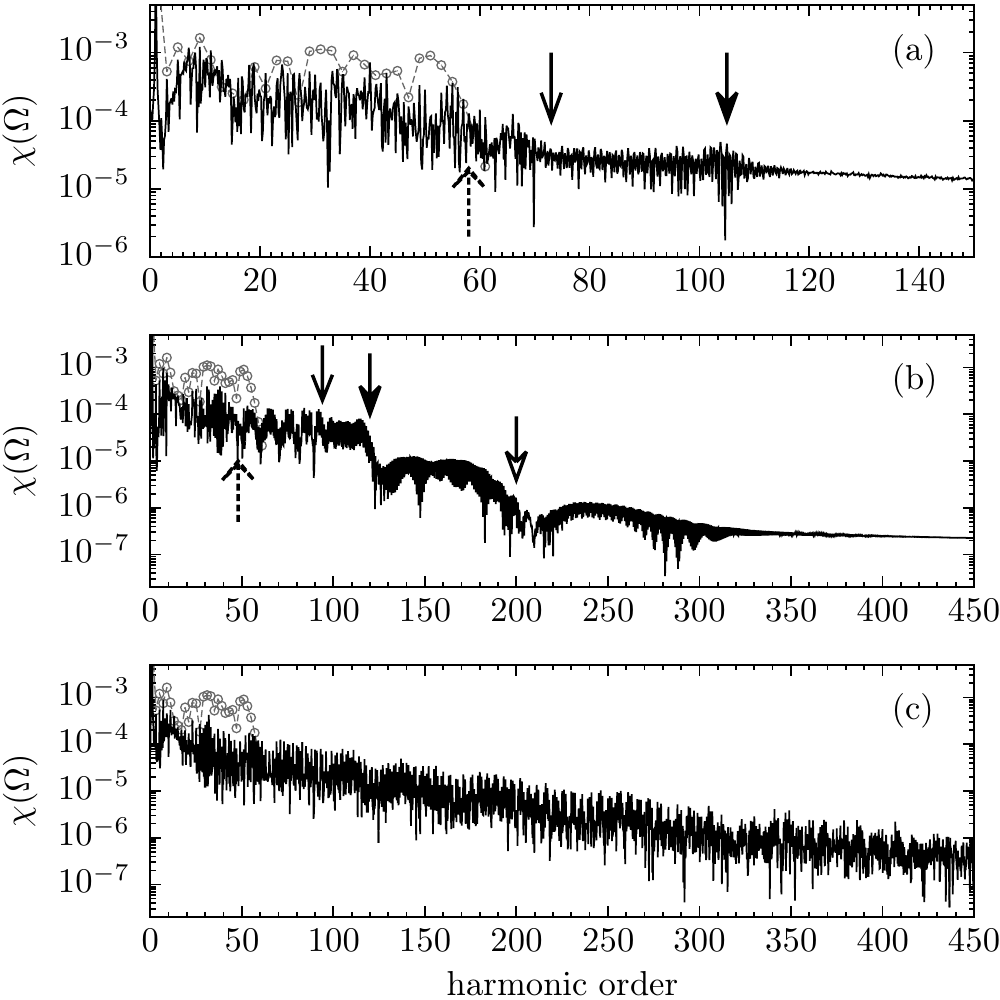}
	\caption{HHG spectra computed from the dipole acceleration (\ref{eq:acceleration}) for different values of the inhomogeneity parameter $\beta$ (solid line) and for the homogeneous case (gray dots). The external field is given by Eq.~(\ref{eq:generalfield}), and its temporal part by Eq.~(\ref{eq:tdfield}), with frequency $\omega=0.05$ a.u., amplitude $E_0=0.075$ a.u. and phase $\phi=-\pi/2$.  The pulse duration is 6 cycles for $\beta=0.002$ and $\beta=0.005$, and 5 cycles for $\beta=0.01$. The cutoff harmonics are indicated by the arrows in the figure. Notice the absence of the cutoff in the panel corresponding to $\beta=0.01$, and that the set of harmonics between 230 and 300 in panel (b) are damped following a ramp-like structure instead of a sharp cutoff. }
	\label{fig:hhgspectra}
\end{figure}

In Fig.~\ref{fig:hhgspectra}, we present the HHG spectra calculated for the system described by the Hamiltonian in (\ref{eq:Hamiltonian}) using several values of $\beta$. These spectra share a series of features, which become more prominent for increasing inhomogeneity parameters. First, the plateau is extended beyond the usual cutoff given by $3.17U_p+I_p$. Second, there are even and odd harmonics, which indicate that the symmetry upon subsequent half cycles of the driving field has been broken. %
Third, the plateau exhibits a  staircase structure, with several cutoffs [see, for instance Fig.~\ref{fig:hhgspectra}(a) and (b)]. This structure becomes more complex as the inhomogeneity parameter increases, until, for around $\beta=0.01$, the spectrum becomes noisy, with no apparent cutoff [Fig.~\ref{fig:hhgspectra}(c)].

\begin{figure}
	\centering
	\includegraphics[width=8cm,angle=0]{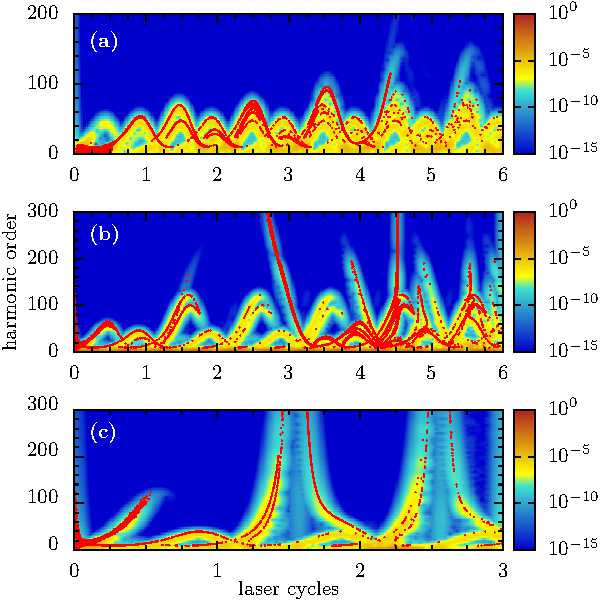}
	\caption{Time-frequency maps and classical returning times (superimposed dots) as functions of the harmonic order and the field cycles, obtained with inhomogeneity parameters $\beta=0.002$ (panel (a)) $\beta=0.005$ (panel (b)) and $\beta=0.01$ [panel (c)]. The kinetic energy of the electron upon return is related to the harmonic frequency by $E_{kin}(t_R)=\Omega-I_p$, where $I_p$ is the ionization potential. The remaining field and atomic parameters are the same as in the previous figure.}
	\label{fig:gabor1}
\end{figure}
Next, we will look more closely at the features in Fig.~\ref{fig:hhgspectra}, and their physical interpretation. With that aim in mind, we calculate the classical emission times for the electron by using an ensemble of classical trajectories, whose initial conditions are sampled according to the Gaussian distribution in (\ref{eq:psi0}) with $x_0=0$ and $p_0=0$. The evolution of this ensemble is given by Newton's equation of motion
\begin{equation}
	\label{eq:newtoneq}
	\frac{\mathrm{d}^2x}{\mathrm{d}t^2}=-2{\cal E}_t\beta\left(x+\frac{1}{2\beta}\right)-\frac{\partial V_a(x)}{\partial x},
\end{equation}
which is obtained by rewriting the system of two first-order differential equations in (\ref{eq:pssystem}). We consider the ionization times to be spread over the first cycle of the driving field, i.e., $0\le t_0 \le 2\pi/\omega$ and take the return condition to be $x(t_R)=0$. 

Figure \ref{fig:gabor1} displays the outcome of these ensemble computations, superimposed to the time-frequency maps calculated with the Gabor transform. Each red dot in the figure corresponds to an electron trajectory that returned to the core according to Eq.~(\ref{eq:newtoneq}). Overall there is a very good agreement between both computations, with several features that will be analyzed below. 
For small and intermediate $\beta$ [Figs.~\ref{fig:gabor1}(a) and (b)], there are many arch-like structures, whose maxima give the cutoff energies. These structures correspond to pairs of electron return times. A very peculiar feature is that some of these arches split, and, in the classical-ensemble computations, we see only a few trajectories returning. The energies of such trajectories are much higher than that of any trajectory returning earlier. For small values of $\beta$, this splitting occurs after several field cycles [see Figs.~\ref{fig:gabor1}(a) and (b)], while for larger $\beta$ it occurs already after a single field cycle [see Fig.~\ref{fig:gabor1}(c)]. In particular, as $\beta$ increases the time-frequency maps become more complex, with many arches still present but starting to break down. This leads to the multiple cutoffs in the spectra, observed in Fig.~\ref{fig:hhgspectra}(b). Finally, for $\beta=0.01$ only the splitting is present, with a set of arches too close to the ionization 
threshold to influence the HHG spectra [Fig.~\ref{fig:gabor1}(c)]. 

\begin{figure}
	\centering
	\includegraphics[width=8cm,angle=0]{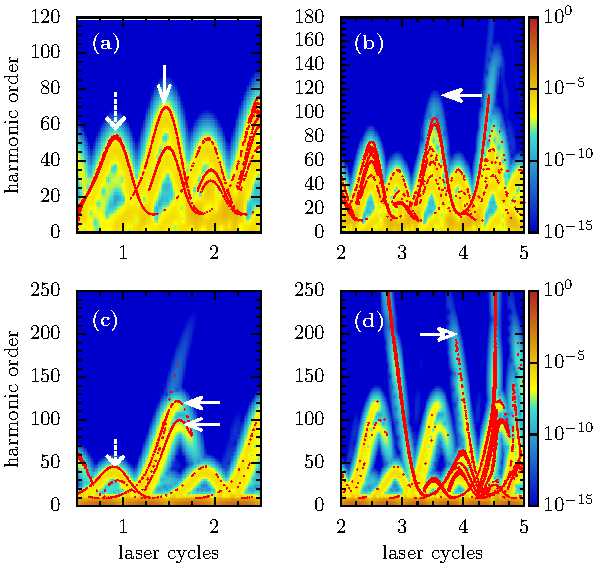}
	\caption{Blow-ups of the time-frequency maps in Fig.~\ref{fig:gabor1}(a) and (b), together with the classical returning times (superimposed dots). Panels (a) and (b) correspond to $\beta=0.002$, while panels (c) and (d) have been computed for $\beta=0.005$. The left and right panels give the Gabor spectra for $0.5T\le t \le 2.5T$ and $2T\le t \le 5T$, respectively, where $T=2\pi/\omega$ is the field cycle. The arrows in the figure, whose styles have been matched with those in Fig.~\ref{fig:hhgspectra}, indicate the cutoff energies.}
	\label{fig:gabor2}
\end{figure}

We will now relate the structures in the Gabor plots to the HHG spectra in Fig.~\ref{fig:hhgspectra}, starting from the shortest, dominant orbits. For these orbits, the electron is released after a field maximum and returns around three quarters of a cycle later, near a crossing of the field. They are widely known as the short and the long orbit \cite{Antoine_1997}, for which the electron returns before and after the field crossing, respectively.  

These orbits are related to the arches plotted in the left panels of Fig.~\ref{fig:gabor2}, in which a blow-up of the first two panels in Fig.~\ref{fig:gabor1} is taken for the interval $0.5T\le t \le 2.5T$. 
In both panels, their maxima differ for subsequent half cycles of the field. This is a consequence of the symmetry breaking introduced by the inhomogeneity, and agrees with the two phase-space configurations in Fig.~\ref{fig:phaseportraits}. 
In particular, around odd numbers of half cycles, i.e., for $t=(2n+1)\pi/\omega$, the cutoff energy increases. This set of times corresponds to the electron being released at a time interval for which ${\cal E}_t<0$, i.e., for the two-saddle configuration. In contrast, there is a decrease in the cutoff energy for arches around full cycles of the field. In this case, ionization occurs at a time interval for which ${\cal E}_t>0$, i.e., when the two centers are present.

The argument put across in \cite{ciappina2012,Ciappina2012b,shaaran2012,shaaran2013,Perez2013} relates this increase to the higher kinetic energy of the electron at the instant of ionization. This is consistent with the slopes in the separatrix observed in the previous section for both configurations, which decrease [increase] for ${\cal E}_t>0$ [${\cal E}_t<0$]. We have verified that these two sets of orbits give the dominant cutoffs in the HHG spectra, marked by the first arrows from the left in Figs.~\ref{fig:hhgspectra}(a) and (b).

Nonetheless, there is evidence that the type of confinement caused by an additional harmonic potential may increase the cutoff energy by bringing trajectories back to the core that otherwise would be irreversibly ionized \cite{FR_2000}. This means that a higher kinetic energy upon ionization would not be the sole mechanism contributing to the energy increase.  One should note, however, that only for the short orbit this additional confinement plays a role. For the long orbit, the electron will return after the crossing, so that $V_l(x,t)$ will be convex. For this reason, as $\beta$ increases the right-hand side of the arches collapse. Physically, this means that the electron may no longer return after the field crossing, i.e., following the long orbit.   
                                                         
The right panels of Fig.~\ref{fig:gabor2} consider the structures in the time-frequency map that develop over longer times, i.e., the splitting leading to very high harmonic energy shown in Fig.~\ref{fig:gabor1}. Throughout, we see that the structures in the time-frequency maps can be linked to specific sets of harmonics in Figs.~\ref{fig:hhgspectra}(a) and (b). For instance, the high-energy arch in Fig.~\ref{fig:gabor2}(b) is related to the low-intensity plateau followed by a cutoff near harmonic order $N=100$ in Fig.~\ref{fig:hhgspectra}(a). Similarly, the high-energy structures in Fig.~\ref{fig:gabor2}(d) lead to the two low-intensity, high-energy branches of the plateau in Fig.~\ref{fig:hhgspectra}(b). A common feature is that, because these structures arise over long time scales, they correspond to electron orbits that spent a long time in the continuum. Hence, there will be a substantial spread of the electronic wave packet, so that its overlap with the ground state upon recombination will be small. 
This leads to low harmonic intensities. 

 Two points regarding the above-mentioned splitting are important to state here. First, the splitting in the arches occurs regardless of how small the inhomogeneity parameter is, provided that the propagation time is sufficiently long. Larger values of $\beta$ only lead to it occurring for shorter times, but do not alter this behavior qualitatively [see Figs.~\ref{fig:gabor1}(c) and (d) for a direct comparison].  Second, such a splitting appears regardless of whether we consider the atomic potential $V_a(x)$ or not.  The features presented in Fig.~\ref{fig:gabor1}, such as the different arch-like structures and the splitting are similar to those in \cite{ciappina2012,Ciappina2012b,shaaran2012}. In these references, however, the splitting has only been identified for relatively large values of $\beta$. 

We will now investigate if the symmetry breaking introduced by the inhomogeneity is responsible for this splitting and the disappearance of the long orbits, as suggested in \cite{shaaran2012}. In order to address this question, we have modified the laser-field potential so that we consider only one of the configurations in Fig.~\ref{fig:phaseportraits}, which then flips with regard to the coordinate $x$ at each half cycle. Explicitly, we employ

\begin{equation}
\widetilde{V}^{(\pm)}_l(x,t)=\pm E_0 |\sin \omega t|
\left[\beta\left(x+\frac{(-1)^n}{2\beta}\right)^2-\frac{1}{4\beta}\right] ,
\label{eq:flipping}
\end{equation}
where $n$ is an integer that increases with the number of half cycles, and has been chosen so that in the first half cycle, $n=0$. $\widetilde{V}^{(+)}_l(x,t)$ is related to the two-center configuration identified in Eq. (\ref{eq:generalfield}) for ${\cal E}_t>0$, while $\widetilde{V}^{(-)}_l(x,t)$ gives the additional saddle. One should note that $\widetilde{V}^{(\pm)}_l(x,t)=\widetilde{V}^{(\pm)}_l(-x,t+\pi/\omega)$, so that the inversion symmetry for subsequent half cycles of the field is not broken. This property also holds for Eq.~(\ref{eq:vlaser}) if $\beta=0$.

\begin{figure}
\centering
\includegraphics[width=9cm,angle=0]{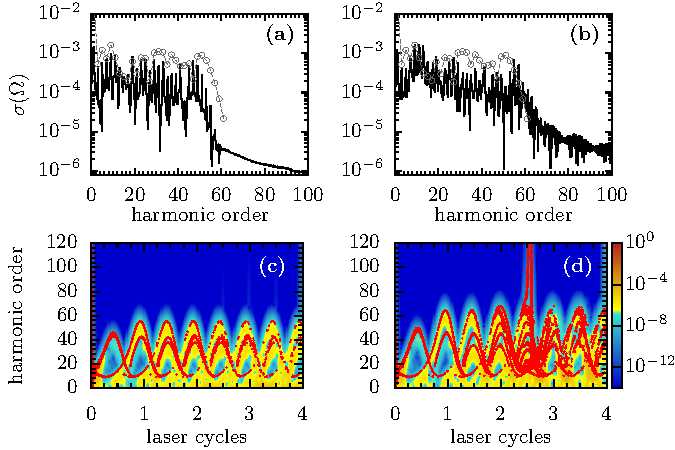}
\caption{Fourier spectra and time frequency maps [upper and lower panels, respectively] computed for $\beta=0.002$, and the same field parameters in Fig.~\ref{fig:hhgspectra}, but using the two symmetric potentials defined in~\ref{eq:flipping}. Panels (a) and (c) correspond to the two-saddle potential $\widetilde{V}^{(-)}_l(x,t)$, while panels (b) and (d) have been calculated using the two-center potential $\widetilde{V}^{(+)}_l(x,t)$. }
\label{fig:flip}
\end{figure}

In Fig.~\ref{fig:flip}, we show the spectra obtained for the two configurations, together with the time frequency maps and classical-trajectory computations [upper and lower panels, respectively]. Both spectra show only odd harmonics, due to the symmetry of $ \widetilde{V}^{(\pm)}_l(x,t)$ for subsequent half cycles. However, only for the two-center configuration there is a visible increase in the cutoff energy, together with several oscillations beyond the cutoff [see Fig.~\ref{fig:flip}(a) in comparison with Fig.~\ref{fig:flip}(b)]. 

These features are consistent with the time-frequency maps and classical-trajectory computations. Once more the trajectories match the time-frequency maps, but the symmetry for subsequent half cycles is no longer broken. The increase in the cutoff can be seen in the arches of Fig.~\ref{fig:flip}(d), and is related to the additional confinement provided by a concave potential. This potential forces trajectories back to the core that would otherwise be irreversibly ionized.  This effect outweighs the smaller momentum at ionization for this type of configuration, and has been discussed in \cite{FR_2000} for a static confining potential. For the same reason, there is no loss of harmonic intensity when the cutoff is extended, as seen in Fig.~\ref{fig:flip}(b). We have also verified that larger values of $\beta$ no longer blocks the electron's return along the long orbit for the dominant pair of orbits. In fact, for $ \widetilde{V}^{(+)}_l(x,t)$, the arches degrade and split, but this splitting is symmetric around 
the cutoff. This confirms that the change in the phase-space configuration after a field crossing is responsible for the removal of the long orbit.

 Notably, only if the additional center is present does one observe the splitting in the arch-like structures. If the two saddles are present, the electron picks up more momentum upon ionization, but it is more difficult for it to return and the high-frequency structures do not arise. This suggests that (i) these structures are directly related to the electronic wave packet being trapped in the additional center around $x_{\beta}$, and that (ii) they are independent of the long-orbit suppression that occurs for shorter time scales. The splitting leads to a second, much lower plateau extending beyond the 120$^{\mathrm{th}}$ harmonic.

\subsection{Analytical model and connection with Mathieu's equation}
\label{subsec:analytical}

 The discussions in the previous section suggest that a model which neglects the core and is restricted to small values of the inhomogeneity parameter suffices for our purposes. Even more, by neglecting the atomic interaction and changing the variables to $Q= 2\beta x+1$ and $\tau=\omega t$, Eq.~(\ref{eq:newtoneq}) can be written in the form of a Mathieu's equation
\begin{equation}
\frac{\textrm{d}^2 Q}{\textrm{d}\tau^2}+\epsilon Q\cos\tau=0,
\label{eq:mathieueq}
\end{equation}
where $\epsilon=2\beta E_0/\omega^2$. 
This equation has been extensively used to study ion traps \cite{Bluemel1989,Brkic2006,March1997,Paul1990}, and it is well known that the stability of its solutions depends strongly on the parameter $\epsilon$. 

 In this work, unless otherwise stated, we will restrict our parameters $\beta$, $E_0$ and $\omega$ in such a way that $\epsilon$ lies within the stability region. The stability condition provides an upper bound for $\beta$ according to 
 \begin{equation}
 0<2\beta \alpha\le 0.439,
 \end{equation}
where $\alpha=E_0/\omega^2$ is the electron's excursion amplitude.  The binding potential, which has been omitted in this model, provides additional confinement in some cases and contributes to extending this region. 

 Fig.~\ref{fig:hhgspectra} shows how this stability condition affects the HHG spectra. The two upper panels have been computed within the stability region, while Fig.~\ref{fig:hhgspectra}(c) is a borderline case. The spectra in  Figs.~\ref{fig:hhgspectra}(a) and (b) exhibit clear plateau and cutoff structures, while the spectrum in Fig.~\ref{fig:hhgspectra}(c) is much noisier, with no clear cutoff.  These features are caused by longer orbits, and have physical, rather than numerical origin. Similar spectra have been found in \cite{Ciappina2012b} for inhomogeneity parameters larger or equal to $\beta=0.01$. In this parameter range, the only way to observe a cutoff was to eliminate the longer trajectories by introducing smaller grids. Therein, a splitting has also been identified in the time-frequency maps for very short time scales. 

\begin{figure*}
\centering
\includegraphics[width=11cm,angle=0]{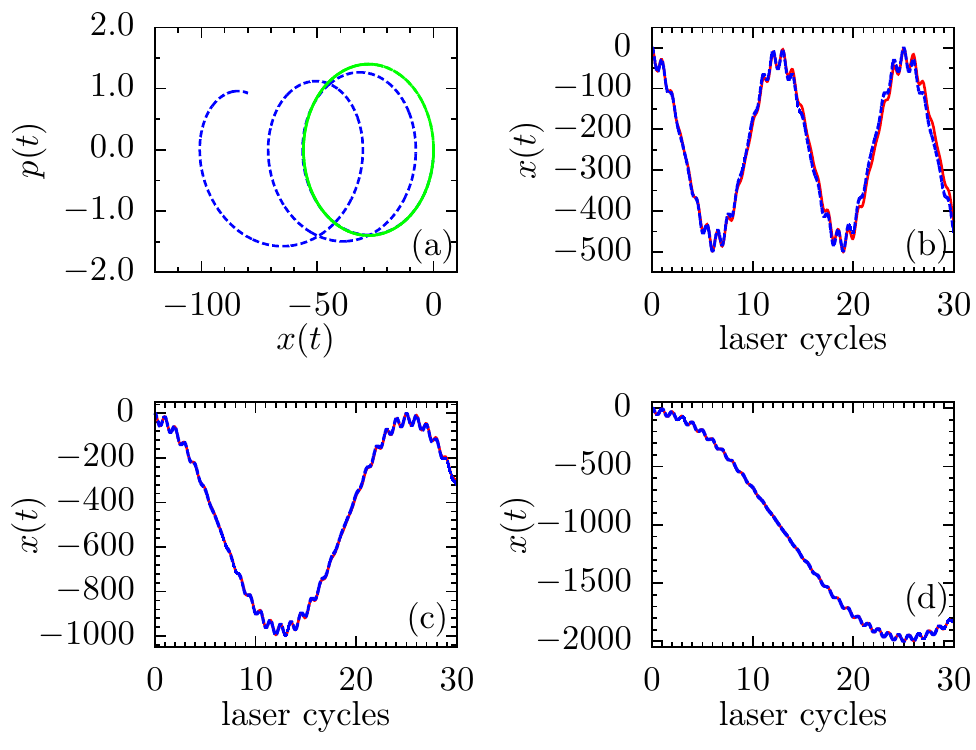}
\caption{Numerical solutions of Mathieu's equation (dashed line) in phase space (panel (a)) and as a position-time plot for $\beta=0.002$ (panel (b)), $\beta=0.001$ (panel (c)) and $\beta=0.0005$ (panel (d)). The initial positions and momenta in the three cases is $x(0)=0$ and $p(0)=0$, respectively. For reference, the continuous line in panel (a) shows a closed orbit resulting from the propagation under a homogeneous laser field. Continuous lines in panels (b), (c) and (d) represent Dehmelt's approximation to Mathieu's equation.}
\label{fig:orbits}
\end{figure*}

In order to provide an explanation in terms of classical concepts, it is useful to look at the behavior of individual trajectories either in phase space or in a position-time plot. This is shown in Fig.~\ref{fig:orbits}, where we can observe two main features. First, in contrast to the homogeneous case, there are no longer closed orbits of period $T=2\pi/\omega$. Instead, we observe that, at each period of time $T$, the orbit becomes displaced from its position at $t=0$ (panel (a) in Figure \ref{fig:orbits}). Second, regardless of the initial conditions and the inhomogeneity parameter, the trajectories experience two kinds of motion, namely one with a small and rapid oscillation and another one with a large and rather slow oscillation (panels (b), (c) and (d) in Figure \ref{fig:orbits}). Moreover, the amplitude and frequency of the large oscillation depend on the inhomogeneity parameter. Both the period and the amplitude of the large oscillation decrease for increasing values of $\beta$.  

The above-mentioned features suggest that the Dehmelt approximation is applicable 
to Mathieu's equation within this range of $\beta$ values. The approximation consists in assuming that the solution can be expressed as a superposition of two motions, one oscillating with a rapid frequency and the other oscillating with a low frequency. This gives
\begin{equation}
\label{eq:superposition}
Q(t)\approx R(t)+L(t),
\end{equation}
where $R(t)$ plays the role of the rapid and small-amplitude oscillation and $L(t)$ that of the slow and large-amplitude oscillation. Notice that $t$ is in atomic units of time. In addition to that, we assume that the high-frequency oscillation amplitude is much smaller than that of the low-frequency motion, which implies that the behavior of $L(t)$ does not affect that of $R(t)$. By doing so, we arrive at an approximation for $R(t)$ and $L(t)$ which read as
\begin{align}
R(t)&\approx\frac{2\beta E_0}{\omega^2} L(t)\cos \omega t\, , \\
L(t)&\approx A\cos(\Omega_s t+\phi_0)\, ,
\end{align}
which allows us to write an approximation for $Q(t)$ as
\begin{equation}
Q(t)=\left[\frac{\sqrt{2}\Omega_s}{\omega}\cos(\omega t)+1\right]A\cos(\Omega_s t+\phi_0)
\label{eq:dehmeltapp}
\end{equation}
where $\Omega_s=\sqrt{2}\beta E_0/\omega$, $\phi_0=-\arctan[\dot Q_0/\Omega_sQ_0]$ and $A=[\Omega_s^{-2}\dot Q_0^2+Q_0^2]^{1/2}/(\sqrt{2}\Omega_s/\omega+1)$, with $\dot Q_0\equiv\mathrm{d}Q/\mathrm{d}t|_{t=0}$ and $Q_0\equiv Q(0)$. For simplicity,  in the above equation, we have considered the initial time of the trajectories to be $t_0=0$. The same line of argument can however be employed for arbitrary ionization times. 

A comparison between the solutions obtained numerically and using Dehmelt's approximation are shown in Fig.~\ref{fig:orbits}. Therein it is clearly observed that the approximation is better suited for small values of the parameter $\beta$. For $\beta=0.001$ and $\beta=0.0005$ we see no difference between the analytical expression and the numerical solution. Furthermore, it is noticeable the fact that although for larger values of $\beta$ discrepancies between the analytical approximation and the numerical solution arise, the period of the slow oscillation given by $T_s=2\pi/\Omega_s$ agrees with the numerical result even for larger values of the inhomogeneity parameter.

Although by using (\ref{eq:dehmeltapp}) the times at which the classical trajectories return to the atomic core can be calculated numerically, an analytical approximation can be derived by using the assumption $\Omega_s \ll \omega$. In order to achieve that, we set $Q(t_R)=1$, which happens when the original variable $x(t_R)=0$. The assumption $\Omega_s \ll \omega$ implies $\sqrt{2}\Omega_s/\omega+1\approx 1$ and $\sqrt{2}\Omega_s\cos(\omega t)/\omega+1\approx 1$. This gives $A\approx [\Omega_s^{-2}\dot Q_0^2+Q_0^2]^{1/2}$ and 
\begin{equation}
1\approx [\Omega_s^{-2}\dot Q_0^2+Q_0^2]^{1/2}\cos(\Omega_st_R+\phi_0)\, ,
\label{eq:dehmeltapp2}
\end{equation}
so that the approximate return times read
\begin{align}
\label{eq:returningtime}
t_R(Q_0,\dot Q_0)=&\frac{1}{\Omega_s}\arccos\left[\frac{\Omega_s}{(\dot Q_0^2+\Omega_s^{2}Q_0^2)^{1/2}}\right]+\frac{2n\pi}{\Omega_s}\\ \nonumber
    &+\frac{1}{\Omega_s}\arctan\left(\frac{\dot Q_0}{\Omega_sQ_0}\right)\, ,
\end{align}
with $n$ an integer (since $\cos$ is a $\mod(2\pi)$-function). This formula gives an approximation to the time that a trajectory takes to return to the core as a function of its initial conditions. For an electron with vanishing position and momentum, $\dot Q_0=0$ and $Q_0=1$. This gives $t_R(1,0)=2n\pi/\Omega_s$, which is the period for the motion shown in Fig.~\ref{fig:orbits}. However, for trajectories with non-vanishing position and momentum, the return times $t_R$ will differ, depending on the initial conditions.

To corroborate that it is indeed the secular oscillation the responsible for the splitting we then compute the momentum by taking the time derivative of (\ref{eq:dehmeltapp}), from which the kinetic energy can be calculated. By inserting the times obtained with (\ref{eq:returningtime}) one can get the kinetic energy at the time when the electron returns to the core. 

\begin{figure}
\centering
\includegraphics[width=8cm,angle=0]{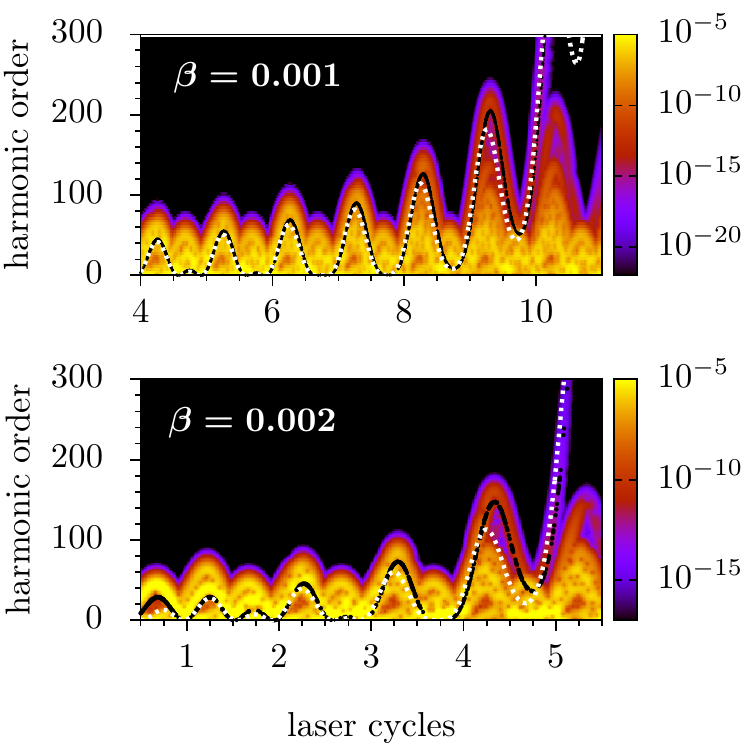}
\caption{Time-frequency maps computed for $\beta=0.001$ (top) and $\beta=0.002$ (bottom), together with the kinetic energies of trajectories returning to the core for $\beta=0.001$. Here, the dotted (white) line indicates the analytical result obtained by inserting (\ref{eq:returningtime}) into the time derivative of Dehmelt's approximation in (\ref{eq:dehmeltapp}) whereas the black dots show the results obtained by evolving the ensemble of classical trajectories placed initially at the core. In both, the analytical approximation and the ensemble of trajectories the initial conditions, at the initial time $t_0=0$, are sampled from a Gaussian distribution with parameter width $\gamma=0.5$.}
\label{fig:analytical_returnings}
\end{figure}

These results are shown in Figure~\ref{fig:analytical_returnings}, together with the Gabor spectra and the classical return times obtained with an ensemble of trajectories. Therein, we can see that, although for  larger values of $\beta$ discrepancies between the model and the ensemble of trajectories result are clearly visible, the splitting in the arches is well predicted and the agreement with the Gabor transform is rather good.
\section{Conclusions}
\label{sec:conclusions}
In this work, we address HHG in spatially inhomogeneous fields, with focus on phase-space and time-scale considerations, using the numerical solution of the time-dependent Schr\"odinger equation (TDSE). We consider reduced-dimensional, one-electron models in which the inhomogeneous field is approximately linear.  This is a widespread assumption, and a good approximation for small values of the inhomogeneity parameter $\beta$ \cite{ciappina2012,Ciappina2012b,shaaran2012,Yavuz_2012,shaaran2013,Hekmatara_2014,Luo_2014}. We find that, in general, the HHG spectra exhibit a rather complex structure, with multiple plateaux and very large cutoff energies. These features are more complex than those reported in the literature, and involve several timescales, which, for small enough $\beta$, can be disentangled using Mathieu's equation and Dehmelt's approximation. 

 All characteristics encountered in the spectra can be traced back to electron trajectories returning to the core using time-frequency analysis and classical-trajectory computations. These trajectories are influenced by two different phase-space configurations that arise for the inhomogeneous field. For subsequent half cycles, the inhomogeneity creates an additional saddle or center, which alters the electron ionization, recombination and propagation in the continuum. This will add to the Stark saddle and to the center located near the minimum of the atomic potential.
 
 For the dominant pairs of orbits, we identify typical arch-like structures whose maxima give the cutoff. These structures are periodic upon a field cycle, and, for an odd number of half cycles of the field, there is an increase in the cutoff energy. Usually, this energy increase is attributed to a higher electron momentum at the instant of ionization. Our results indicate, however, that this is not the only important mechanism. Another key ingredient is the additional confinement introduced by the fact that the potential is concave in the subsequent half cycle of the field. This confinement forces high-energy orbits, which otherwise would be irreversibly ionized, back to the core. Furthermore, because the confining effect is only present up to the subsequent field crossing, it affects the short and the long orbit unequally, until, for larger values of $\beta$, the contributions of the long orbit are suppressed. This brings additional insight on the suppression of the long orbit, which has been identified 
and analyzed in several publications \cite{ciappina2012,Ciappina2012b,shaaran2012,Yavuz_2012,shaaran2013,Hekmatara_2014,Luo_2014}. To be able to control the suppression of the long trajectory is particularly important for the generation of attosecond pulses and of supercontinua, as discussed in \cite{Yavuz_2012,Hekmatara_2014,Luo_2014}.
 
 This has been exemplified by constructing two effective laser-interaction potentials $\widetilde{V}^{(\pm)}_l(x,t)$, which lead to only one of the above-stated configurations and have the same symmetry properties as the homogeneous field. For $\widetilde{V}^{(-)}_l(x,t)$, the electron reaches the continuum with a higher momentum but there is no additional confinement upon return, while for $\widetilde{V}^{(+)}_l(x,t)$ there is a lower momentum at the instant of ionization, but confinement upon return. Only for the latter potential have we found a higher cutoff energy for the two dominant orbits, which indicates that confinement is more important. A similar effect has been identified in our publication \cite{FR_2000}, for which, however, the confining potential was static. For the symmetric potentials, the arch-like structures did not collapse, and the long orbits in the dominant pair were not eliminated. This provides support for our argument that the different configurations around the field crossing lead 
to this effect.

For the longer pairs of trajectories, we identify a splitting in the arch-like structures, which leads to extremely high harmonic frequencies.  This splitting leads to further, much lower plateaux in the spectra. For low enough values of $\beta$, we determine the times for which the splittings occur analytically using Mathieu's equation and Dehmelt's approximation. We also have found evidence that confinement is important in order to obtain such structures, as they are absent for the auxiliary potential $\widetilde{V}^{(-)}_l(x,t)$.

The present studies also invite the following, more speculative questions. First, it seems that two centers, one of which is created by the inhomogeneity, are necessary if the cutoff energy is to be extended. This resembles the case of molecular HHG, for which a substantial cutoff extension has been reported \cite{Kopold_1998}.  Hence, it would be of interest to assess whether similar effects to those reported here could be seen in molecules. Second, schemes to increase the harmonic efficiency in the energy regions related to the above-mentioned splitting would be very desirable, as they would provide us with extremely high-frequency sources. This may be possible to achieve by modifying the geometry of the nanostructures producing the field, or by an appropriate choice of macroscopic propagation conditions. 

Finally, one should note that the extension of the cutoff and/or the collapse of the long orbits are not specific to the linear spatial inhomogeneity studied here. In fact, these features have been reported for fields of other functional forms such as that employed in \cite{shaaran2013}. Thus, the features studied in the present work may contribute to a more general insight in plasmonically enhanced HHG.

We would like to thank T. Shaaran, M. Lewenstein and N. Braz for useful discussions and S. Rai for providing references. This work has been funded by the UK EPSRC (grant EP/J019240/1 and summer bursary).

\end{document}